\documentclass[pra,aps,twocolumn,superscriptaddress,longbibliography,floatfix,notitlepage]{revtex4-1}
\usepackage{dcolumn}%
\usepackage{bm}%
\usepackage[english]{babel}
\usepackage[T1]{fontenc}
\usepackage{lmodern}
\usepackage{amsmath,amsfonts,amssymb,amsthm,bm,times,dcolumn}
\usepackage{cancel}
\usepackage{mwe}
\usepackage{microtype}
\usepackage{braket}
\usepackage{mathrsfs}
\usepackage{nicematrix}
\usepackage{array}
\usepackage{tikz, circuitikz}
\usetikzlibrary{arrows.meta, positioning, decorations.pathmorphing}
\usepackage{pgfplots}
\usepgfplotslibrary{colormaps,patchplots}
\pgfplotsset{compat=1.18}
\usepackage{blkarray}
\usepackage[colorlinks={true},citecolor={blue},filecolor={blue},linkcolor={blue},urlcolor={blue}]{hyperref}
\usepackage{graphicx, epstopdf, color}
\usepackage{subfigure}
\usepackage[mathscr]{euscript}
\usepackage{setspace}
\usepackage{physics}
\usepackage{stmaryrd}

\newcommand{\kb}[2]{\ket{#1}\bra{#2}}

\newcommand{\idg}[1]{{\bfseries #1)}}

\newcommand{\subfigimg}[3][,]{%
	\setbox1=\hbox{\includegraphics[#1]{#3}}%
	\leavevmode\rlap{\usebox1}%
	\rlap{\hspace*{2pt}\raisebox{\dimexpr\ht1-0.5\baselineskip}{{\bfseries \large\textsf{#2}}}}%
	\phantom{\usebox1}%
}

\begin{document}

\title{Noise-Resilient Quantum Evolution in Open Systems through Error-Correcting Frameworks}

\author{Nirupam Basak}
\email{nirupambasak2020@iitkalumni.org}
\affiliation{Indian Statistical Institute, Kolkata 700108, India}

\author{Goutam Paul}
\email{goutam.paul@isical.ac.in}
\affiliation{Indian Statistical Institute, Kolkata 700108, India}

\author{Pritam Chattopadhyay}
\email{pritam.chattopadhyay@weizmann.ac.il}
\affiliation{Weizmann Institute of Science, Rehovot 7610001, Israel}

\date{\today}%

\begin{abstract}
We analyze quantum state preservation in open quantum systems using quantum error-correcting (QEC) codes explicitly embedded in microscopic system-bath models. Rather than assuming abstract quantum channels, we consider multi-qubit registers coupled to bosonic thermal environments, derive a second-order master equation for the reduced dynamics, and use it to benchmark the five-qubit, Steane, and toric codes under local and collective noise. We compute state fidelities as functions of system-bath coupling strength, bath temperatures, and the number of correction cycles. In the low-temperature regime, repeated error correction with the five-qubit code significantly suppresses decoherence and relaxation for weak-to-moderate couplings. In the high-temperature regime, thermal excitations reduce the effectiveness of all codes, although within the parameter range studied, the five-qubit code still yields the highest fidelities among the three codes. For two-qubit Werner states, we identify a critical evolution time associated with an early-time crossover, before which the overhead of QEC does not compensate for the noise-induced degradation; this critical time increases with entanglement, reflecting the greater fragility of strongly entangled states. Overall, our results provide a microscopic master-equation-based framework for benchmarking QEC performance in realistic open-system environments and for assessing code behavior in near-term noisy quantum architectures.
\end{abstract}

\maketitle

\section{Introduction}\label{Intro}

Quantum computing (QC) technologies have rapidly evolved, heralding transformative potential across computational and information-processing domains~\cite{grover1996fast,shor1999polynomial, arora2009computational, nielsen2010quantum,gebhart2021quantifying}. The theoretical promise of quantum speedup is now being substantiated through practical demonstrations on first-generation \textit{quantum processors}~\cite{arute2019quantum, zhong2021phase}. However, transitioning from these early-stage \textit{noisy intermediate-scale quantum} (NISQ) devices~\cite{knill2005quantum,preskill2018quantum,georgopoulos2021modeling,xie2025advances} to scalable, fault-tolerant quantum architectures remains an immense challenge. At the heart of this challenge lies the urgent need to mitigate the detrimental impact of environmental noise and system imperfections.  It undermines the delicate quantum superpositions required for the reliable operation in quantum computing processes~\cite{steane1998quantum,nielsen2010quantum}, reduces the fidelity of quantum communication protocols~\cite{kremer1995quantum,nielsen2010quantum}, and severely limits the efficacy of quantum memories and sensing~\cite{heshami2016quantum,gsh7r7ms,Correa2015,PritamQST2,Mehboudi2019,Potts2019,PRXQuantum.5.030338}. Both quantum error correction (QEC)~\cite{shor1995scheme, steane1996error, gottesman1996class, laflamme1996perfect, steane1996error, calderbank1996good, steane1996multiple, bennett1996mixed, knill1997theory, gottesman1997stabilizer, chuang1997bosonic, calderbank1997quantum, calderbank1997quantum, knill1998resilient, cory1998experimental, braunstein1998quantum, cochrane1999macroscopically, kribs2005unified, aliferis2005quantum, oreshkov2007continuous, aliferis2008accuracy, fowler2009high, ng2010simple, wang2011surface, mandayam2012towards, fowler2012surface, reed2012realization, yao2012experimental, lidar2013quantum, criger2016noise, michael2016new, ofek2016extending, loong2018open, babu2023quantum, tanggara2024strategic, tanggara2024simple, basak2025approximate, basak2025resource, basak2025multiparty, spencer2025qudit, Fu2025errorcorrectionin} and fault tolerance~\cite{aharonov1997fault, kitaev2003fault, steane2003overhead, terhal2005fault, aharonov2006fault, raussendorf2007fault, aharonov2008fault, ng2009fault, stephens2014fault, hillmann2025singleshot, dutkiewicz2025error, akahoshi2025compilation} are essential pillars for the reliable execution of quantum information processing (QIP) tasks, but they demand sophisticated theoretical frameworks and precise experimental implementation.

QEC, introduced independently by Shor~\cite{shor1995scheme} and Steane~\cite{steane1996error}, provides a structured methodology to safeguard quantum information from \textit{decoherence} and other forms of noise-induced degradation~\cite{peres1985reversible, shor1995scheme, bennett1996mixed, laflamme1996perfect, steane1996multiple, knill1997theory, knill1997theory, gottesman1997stabilizer, bravyi1998quantum, cochrane1999macroscopically, lidar2013decoherence, michael2016new, guardia2020quantum}.  
Despite formidable hurdles, several successful realizations of QEC codes have been demonstrated using current quantum technologies~\cite{reed2012realization, yao2012experimental,chattopadhyay2025understanding1, erhard2021entangling, chattopadhyay2025landauer}.
Traditionally, QEC strategies are analyzed using abstract quantum channel models, which encapsulate noise effects without explicitly modeling the underlying physical interactions. However, these channels often emerge as approximations to more fundamental Hamiltonian noise models, where the quantum system is coupled to an inaccessible bath, and the dynamics are governed by system-bath interactions~\cite{breuer2007theory}. While quantum circuits dominated by Markovian noise exhibit higher error thresholds and tractable correction strategies, more general, non-Markovian noise scenarios remain challenging. This raises a crucial question: \textit{Are the simplified noise models typically used in QEC studies adequate for describing realistic, physically grounded quantum noise?}

This work analyzes quantum state preservation in open quantum systems (OQSs) using realistic, physically motivated system–bath models, going beyond idealized quantum-channel descriptions. We explicitly incorporate microscopic system–environment interactions and systematically benchmark the performance of QEC codes, namely, the five-qubit, Steane, and Toric codes, for different system–bath coupling strengths and bath temperatures.  We observe that
(i) in the low-temperature regime, the five-qubit code provides a advantage over others in preserving state fidelity. 
(ii) Repeated error-correction cycles efficiently suppress both decoherence and energy relaxation for weak and moderately strong couplings. 
(iii) In the high-temperature regime, we observe that thermal excitations dominate the dynamics and reduce the overall effectiveness of QEC codes.
(iv) The five-qubit code provides better fidelity over the Steane and toric codes. 
(vi) Extending the analysis to two logical qubits, specifically two-qubit Werner states, we identify a critical evolution time before which QEC fails to improve fidelity. This critical time increases with the degree of entanglement, indicating enhanced fragility of strongly entangled states.

Collectively, these results highlight the interplay between system–bath correlations, thermal noise, and code structure, and establish a quantitative framework for benchmarking QEC performance in realistic open-system environments, providing guidance for the design of noise-resilient quantum architectures.

\section{Many-body system model}\label{Model}
\begin{figure}[htpb]
\centering
\includegraphics[width=\columnwidth]{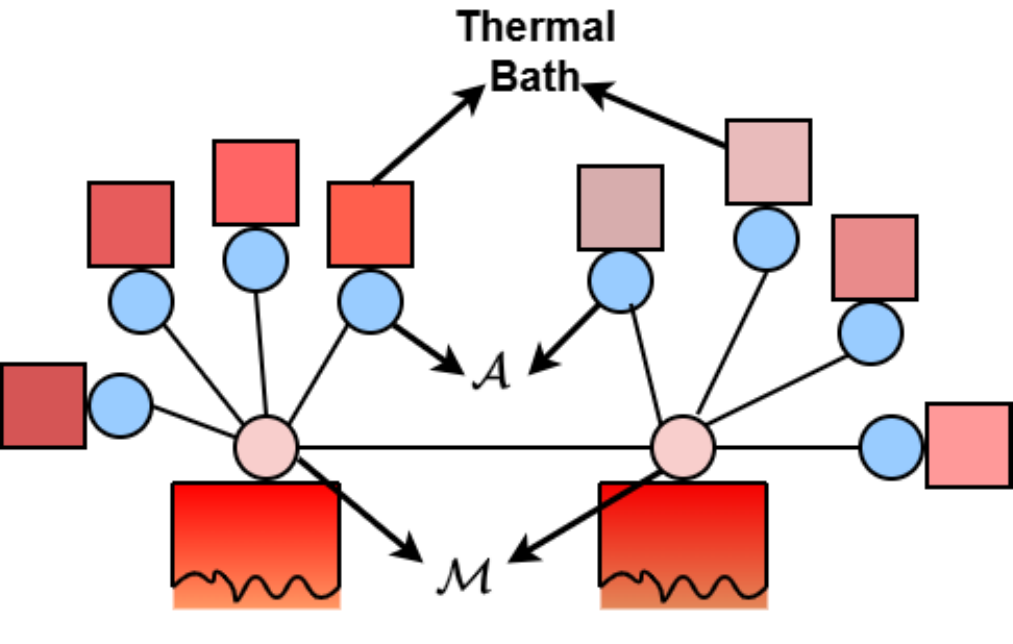}
\caption{Schematic of the system-ancilla and its respective thermal baths. $\mathcal{A}$ denotes the ancilla for the respective main qubit $\mathcal{M}$. The ancilla and the system qubit interact with local thermal baths with different temperatures. The different shades of red are used to designate the different temperatures of the thermal bath.}
\label{fig:system1}
\end{figure}

We consider an open quantum system comprising a two-qubit logical register and a variable number of ancilla qubits, the latter depending on the specific QEC code under investigation (e.g., five-qubit, Steane, or Toric code). The logical and ancilla qubits together form what we denote as the total system. Each qubit in this system interacts with its own independent local/collective environment (Fig.~\ref{fig:system1}), modeled as a bosonic bath, thereby introducing decoherence and dissipation characteristic of OQSs.

The complete Hamiltonian governing the \textit{system-bath dynamics}~\cite{breuer2007theory,rivas2012open} is expressed as a sum over local contributions from each qubit:
\begin{eqnarray}
 H = \sum_j H^{S}_j + H^B_{j} + H^{SB}_{j},  
\end{eqnarray}
where the index $j$ runs over all qubits (both logical and ancillary) in the total system.
Each qubit (system-ancilla) $j$ evolves under a local Hamiltonian $H^{S}_j$ ($S=\mathcal{M}$-$\mathcal{A}$, $\mathcal{M}$ and $\mathcal{A}$ denotes the main system and ancilla qubits, respectively.):  
\begin{eqnarray}
    H^{S}_j = \frac{\omega_j}{2} \hat{\sigma}_{zj},
\end{eqnarray}
where $\omega_j$ is the frequency of the qubit $j$, $\hat{\sigma}_{\gamma j}$ ($\gamma =x, y, z$) are the Pauli operators.  This term encapsulates the free evolution of individual qubits under their respective quantization axes and sets the energy scale for their dynamics in the absence of interactions.

The bath Hamiltonian $H^B_j$ reads as
\begin{eqnarray}
    H^B=\sum_jH_j^B = \sum_j\sum_k\Omega_{k,j}b_{k,j}^\dagger b_{k,j}, 
\end{eqnarray}
where $\Omega_{k,j}$ denotes the frequency of the $k$th mode of the $j$th bath, $b_{k,j}^\dagger$  and $b_{k,j}$ represents the creation and the annihilation operators respectively. The bath is initially in a thermal equilibrium state: $\rho_B = e^{-\beta H^B}/Z$, where $\beta$ and $Z$ characterize the inverse temperature and partition function of the bath.

The bosonic bath weakly interacts with system-ancilla spins. The coupling Hamiltonian yields as
\begin{eqnarray}
    H^{SB}_j=\sum_kg_{k,j}\left(\sigma_j^+b_{k,j}+\sigma_j^-b_{k,j}^\dagger\right),
\end{eqnarray}
where $g_{k,j}$ denotes the coupling strength of the $j$th qubit to the bosonic bath modes. The spin ladder operators $\sigma_{j}^\pm$ correspond to the qubit raising and lowering operators, facilitating energy exchange between the system and the reservoir.

Here, we assume that the system qubits and the ancilla are coupled during the time evolution. Therefore, the dynamics of the system-ancilla are described using the master equation method.

\subsection*{Bath-driven evolution of the system–ancilla}
To describe the dynamics of an OQS under weak coupling to its environment, we employ a time-nonlocal master equation for the reduced density matrix $\rho_S(t)$ of the system. The full system-bath interaction is treated perturbatively up to second order in the interaction Hamiltonian $H_{SB}$, which is valid in the Born approximation and the initial total state is assumed to be factorized. Throughout the analysis we will restrict ourselves to \textit{weak–to–moderate coupling}, with $\kappa/\omega \leq 0.1$, so that the second-order Born approximation remains qualitatively valid; $\kappa/\omega = 0.01 $ and $\kappa/\omega = 0.1 $ are used to represent relatively weaker and moderate couplings within this regime. The resulting integro-differential master equation reads: 
\begin{eqnarray}\label{state}
  \dot{\rho_S}(t) && = -i[H_S(t), \rho_S(t)] \nonumber\\
 &&+ \int^t_{t' = 0} dt'  \Phi(t-t')  [S(t) \rho_S(t), S(t')] + h.c. \nonumber\\
 \end{eqnarray}
$\Phi(t-t')$ is the bath correlation function encapsulating the memory effects of the environment, $H_S(t)$ is the system Hamiltonian, and $S(t)$ denotes the system operator coupled to the bosonic bath in the interaction picture. Eq.~\eqref{state} emerges from tracing out the bath degrees of freedom under the assumption that the bath remains in thermal equilibrium throughout the evolution.

In the interaction picture, one can derive the density matrix of the system, $\rho_S(t)$,  under the assumption of the weak system–bath interaction, to second order in $H_{SB}$~\cite{breuer2007theory}.
In the interaction picture, the coupling Hamiltonian becomes

\begin{eqnarray}
H^I_{SB}(t) = \sum^2_{\alpha =1} {S}_{\alpha}(t) \otimes {B}^{\dagger}_{\alpha}(t), 
\end{eqnarray}
with ${S}_{\alpha}(t) = e^{i H_S t} \tilde{S}_{\alpha} e^{-i H_S t}$ (where $\tilde{S}_{1(2)} = \sum_{i} \sigma^{+(-)}_i$) and ${B}_{\alpha}(t) = e^{i H_B t} \tilde{B}_{\alpha} e^{-i H_B t}$ (where $\tilde{B}_{1(2)} = \sum_k g_k \sigma^{-(+)}_k$). The couplings considered here are $\tilde{B}_1 =  \sum_k g_{kj} b^\dagger_{kj}$ and $\tilde{B}_2 = \tilde{B}^\dagger_1$. 
The bath response functions $\Phi_{\alpha \alpha'}(\tau) = Tr_B(B_\alpha(\tau)B_{\alpha'}(0)\rho_B)$ are shown to be of the following forms (see Appendix~\ref{appendix:A}) 

\begin{eqnarray} \label{Bathfunction}
&&\Phi_{11}(\tau) =  \Phi_{22}(\tau) = 0,\nonumber\\
&&\Phi_{12}(\tau) = \Gamma \big\langle e^{2i\omega \tau} (n(\omega) +1) \big\rangle, \Phi_{21}(\tau) =\Gamma \big\langle e^{-2i\omega \tau} n(\omega) \big\rangle, \nonumber\\ 
\end{eqnarray}
with $\Gamma = \sum_{k} g^2_{k,j}$, a positive prefactor that depends on the nature of the interaction between the bath and the system, and $n(\omega)$ is the mean occupation number of the bath mode at frequency $\omega$. Eq.~\eqref{Bathfunction} reflects energy exchange between the system and the environment.

The time-evolved system operators appearing in the master equation are explicitly:

 \begin{eqnarray}\label{eqnSS}
 \allowdisplaybreaks
&& S_{1}(t) = U^{\dagger}_S (t) \Big( \sum^N_{i=1} \sigma^+_i \Big)  U_S (t),\nonumber \\
&& S_{2}(t) = U^{\dagger}_S (t) \Big( \sum^N_{i=1} \sigma^-_i \Big)  U_S (t),
\end{eqnarray}
where $N$ is the number of main qubits and $U_S (t)= e^{-iH_{S}t}$ is the unitary time evolution operator generated by the system Hamiltonian. The operators $S_\alpha(t)$ capture the collective raising and lowering transitions of the system qubits under time evolution, and their interactions with the bath are responsible for decoherence and population dynamics within the system (Appendix~\ref{appendix:B}).

The described reduced dynamics retain memory effects through the time-nonlocal bath correlation functions, and therefore capture non-Markovian features within each evolution interval. However, while implementing repeated QEC cycles, we will adopt a cycle-wise resetting approximation: after each recovery step, the system is reinitialized in the code space and the bath is assumed to remain in its stationary thermal state. Consequently, system–bath correlations generated during one interval are not propagated to subsequent cycles.

This prescription is therefore intermediate in nature, i.e., it is non-Markovian within each evolution segment, but does not include long-time memory accumulation across multiple QEC cycles.

Unlike the phenomenological error models commonly used in QEC studies, the microscopic system–bath Hamiltonian that we consider here, in principle, generates unknown error processes. However, within the second-order master equation formalism and the weak-to-moderate coupling regime explored, the dynamics are dominated by dissipative contributions arising from bath correlation functions, which effectively lead to incoherent decoherence and relaxation. Coherent contributions, such as Lamb-shift-type renormalizations or phase-coherent error accumulation, are not treated explicitly and are expected to play a subleading role in the parameter regime considered. Moreover, the cycle-wise resetting approximation suppresses the accumulation of coherent errors across successive QEC cycles.

\section{Quantum Error Correcting codes}\label{QEC}
\begin{figure*}[htpb]
\centering
\includegraphics[width=0.9\textwidth]{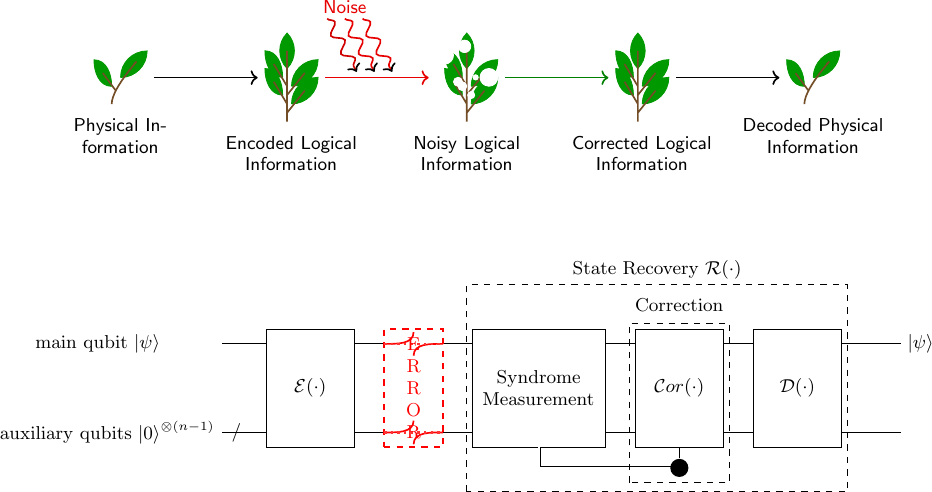}
\caption{\textbf{(Top)} Conceptual overview of the QEC process. Physical information is encoded into logical states before the noise acts and distorts them. After passing through a correction stage, the recovered logical information is finally decoded to reproduce the original physical information. \textbf{(Bottom)} A generic block diagram for $n$-qubit quantum error correcting code. A physical qubit with state $\ket\psi$ is encoded via $\mathcal{E(\cdot)}$ with $n-1$ auxiliary qubits initialized at state $\ket0$ to build one logical qubit, which undergoes some error channel, denoted in red. Then the original state is recovered via some recovery operation $\mathcal{R}(\cdot)$, which consists of syndrome measurement, correction operation $\mathcal{C}or(\cdot)$ and decoder $\mathcal{D}(\cdot)$. By tracing out the auxiliary qubits, one can get back the original state $\ket{\psi}$.}
\label{fig:enter-label}
\end{figure*}

To mitigate the errors caused by the baths, we employ different QEC codes to estimate the success rate of the state restoration to its initial form. For the performance comparison, we focus on the minimal code overhead and fewer physical resources. Therefore, we choose the five-qubit code as the smallest perfect code~\cite{laflamme1996perfect}, the Steane code as the smallest CSS code~\cite{steane1996error, steane1996multiple} and $2\times 2$ toric code as the smallest topological code~\cite{kitaev2003fault}. A generic representation of the QEC code is shown in Fig. \ref{fig:enter-label}. 
The details of the codes are described in this section.

\subsection{Five-qubit code}
To mitigate the effects of environmental decoherence, we employ the five-qubit QEC code~\cite{laflamme1996perfect}, which is the smallest perfect code capable of correcting arbitrary single-qubit errors. This code requires only five physical qubits to protect one \textit{logical qubit} from the influence of local quantum noise. Specifically, for each system qubit, four ancillary qubits are introduced to detect and correct any single-qubit error that may occur during the system’s evolution.

A complete error correction cycle in this code consists of three main stages:

Encoding ($\mathcal{E}$): At the beginning of the qubit's time evolution, the main system qubit is coupled with the four ancillary qubits via a unitary encoding operator $U$. This transforms the single-qubit logical state into a five-qubit coupled codeword within the protected code space.

Error Evolution ($E$): The five-qubit system is then exposed to environmental interactions governed by the error channel $E$, which simulates decoherence or noise processes due to coupling with local reservoirs. The code allows for arbitrary single-qubit errors on any of the five qubits.

Syndrome Extraction and Recovery ($\mathcal{R}$): Finally, the recovery operation $\mathcal{R}$ is applied to restore the original logical state based on the extracted error syndrome. This process is implemented through the inverse of the encoding operation, $U^\dagger$, followed by local measurements on the ancillary qubits. Notably, this phase of the code does not require any additional ancillae beyond the original five qubits. The corrective action is performed on the main qubits to reverse any error that may have affected the logical qubit.

Additional procedural and technical specifics regarding the implementation of the code are provided in Appendix~\ref{appendix:C}.

\subsection{Steane code}
We also consider the Calderbank-Shor-Steane (CSS) code~\cite{steane1996error,steane1996multiple}, which encodes logical information into an entangled subspace using two classical linear codes $C_1$ and $C_2$ with $C_2 \subset C_1$. This structure enables independent correction of bit-flip and phase-flip errors.

The logical qubit is encoded into an $\llbracket n,k,d\rrbracket$ stabilizer code space via Clifford circuits. After evolution under the microscopic error model $E$, stabilizer measurements diagnose errors using parity checks derived from $C_1$ and $C_2^\perp$, followed by a classical decoding and Pauli correction. Further implementation details are given in Appendix~\ref{appendix:D}.

\subsection{Toric code}
The toric code~\cite{kitaev2003fault} encodes logical qubits into global degrees of freedom of a two-dimensional lattice with periodic boundary conditions. Qubits reside on the edges of an $L\times L$ lattice ($n=2L^2$), and the stabilizers are generated by star operators $A_v=\prod_i X_i$ and plaquette operators $B_p=\prod_i Z_i$.

The code space is defined by the simultaneous +1 eigenspace of all stabilizers and supports $k=2$ logical qubits, with distance $d=L$. Logical operators correspond to non-contractible loops across the lattice.

A correction cycle involves encoding into the stabilizer subspace, evolution under the microscopic noise model $E$, and syndrome extraction via stabilizer measurements. Errors are identified as anyonic excitations and corrected using a classical decoder (e.g., minimum-weight perfect matching). Logical errors occur only when error chains form nontrivial loops. Implementation details are provided in Appendix~\ref{appendix:E}.

In this work, finite-size realizations are considered, and encoding and recovery for each code are assumed ideal to isolate environmental effects. Performance is quantified via logical fidelity.

\subsection{Fidelity}

The efficacy of all the error correction codes is quantitatively assessed by evaluating the fidelity between the recovered state and the original logical state. Fidelity serves as the key figure of merit in benchmarking the code's capacity to preserve quantum information in the presence of open-system decoherence. Let us consider a quantum channel $\mathcal{C}$ consisting encoding $\mathcal{E}(\bullet)$, evaluation $E(\bullet)$ and recovery $\mathcal{R}(\bullet)$. Suppose $\rho_0$ denotes the initial quantum state prior to the evolution, and let $\rho_t$ denote the state of the system after it has evolved for a time $t$. Then we have
\begin{equation}
\rho_t=\mathcal{C}(\rho_0)=\mathcal{R}(E(\mathcal{E}(\rho_0))).
\end{equation}
The state fidelity~\cite{jozsa1994fidelity, nielsen2010quantum} between $\rho_0$ and $\rho_t$ is defined as
\begin{equation}
\mathcal{F}_{state}(\rho_0, \rho_t)=\left(\tr\sqrt{\sqrt{\rho_0}\rho_t\sqrt{\rho_0}}\right)^2.
\end{equation}

\section{State preservation via quantum error correction }\label{State}

Here, we employ the QEC codes to investigate the preservation of quantum states in open-system environments. For a comprehensive assessment of their performance, we consider two representative configurations of the logical register: one with a single logical qubit ($\mathcal{M}=1$), corresponding to the protection of an isolated quantum state, and another with two logical qubits ($\mathcal{M}=2$), enabling the examination of entanglement preservation and correlated error dynamics. This dual-scenario analysis allows us to elucidate how different QEC codes respond to environmental noise under varying system sizes, coupling strengths, and temperature regimes, thereby offering a unified framework for evaluating the robustness of encoded quantum information in realistic OQSs.

Throughout this work, the encoding and recovery operations associated with QEC cycles are assumed to be ideal and effectively instantaneous, such that no additional noise or time overhead is introduced during the correction process. As a result, the observed improvement in fidelity with increasing numbers of QEC cycles should be interpreted as an upper bound on performance. In realistic implementations, finite-duration syndrome extraction and imperfect recovery operations would introduce additional decoherence, which may reduce the net advantage of repeated correction cycles. However, the look-up table-based decoding approach from~\cite{laflamme1996perfect} used in this work is not optimal. The performance may be increased by using an optimal decoder, with the cost of exponential time complexity, making the decoding problem NP-hard~\cite{hsieh2011np, kuo2012hardness, basak2026faster}.

\begin{figure*}[htpb]
\centering
\subfigimg[width=0.49\textwidth]{a)}{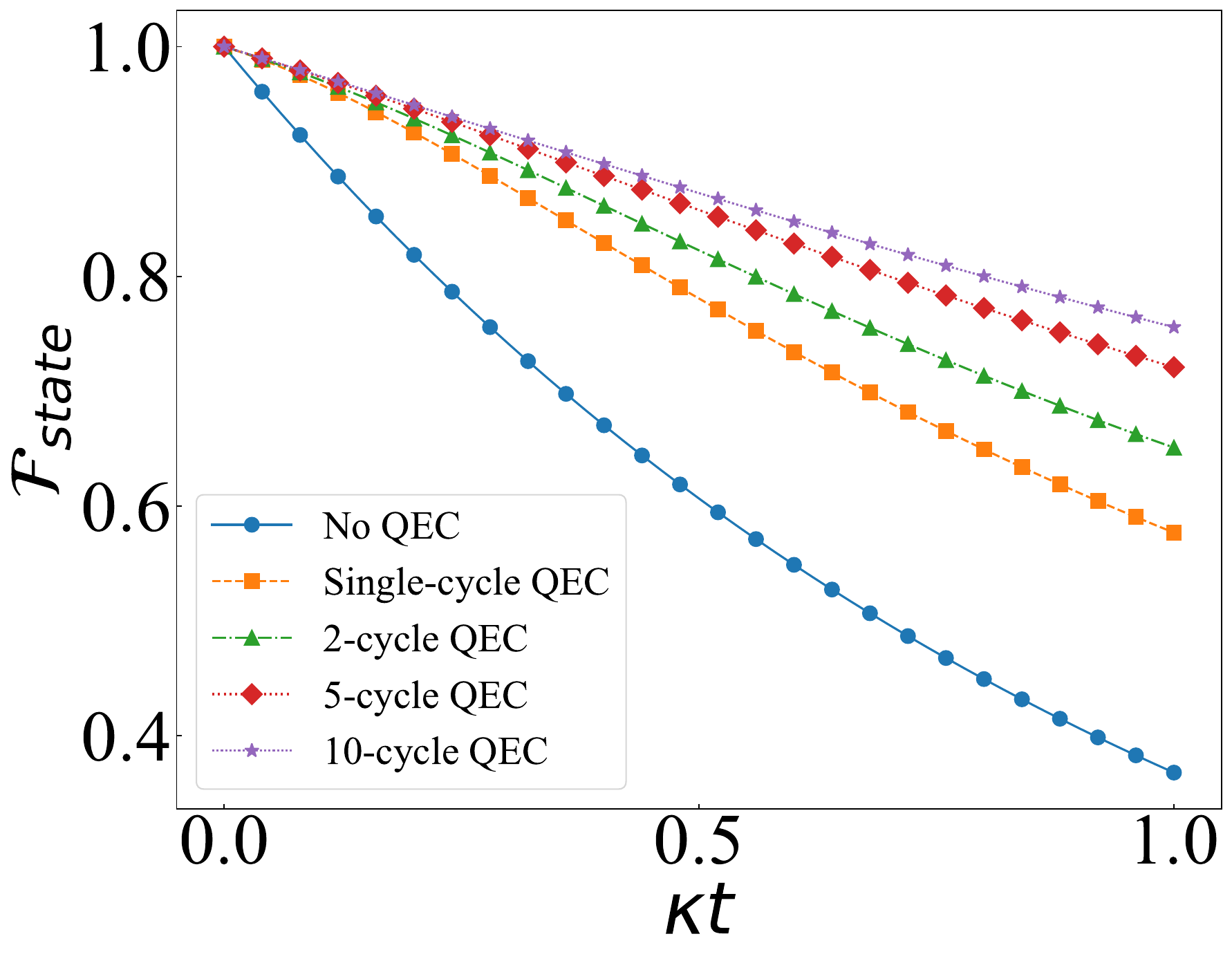}\hfill
\subfigimg[width=0.49\textwidth]{b)}{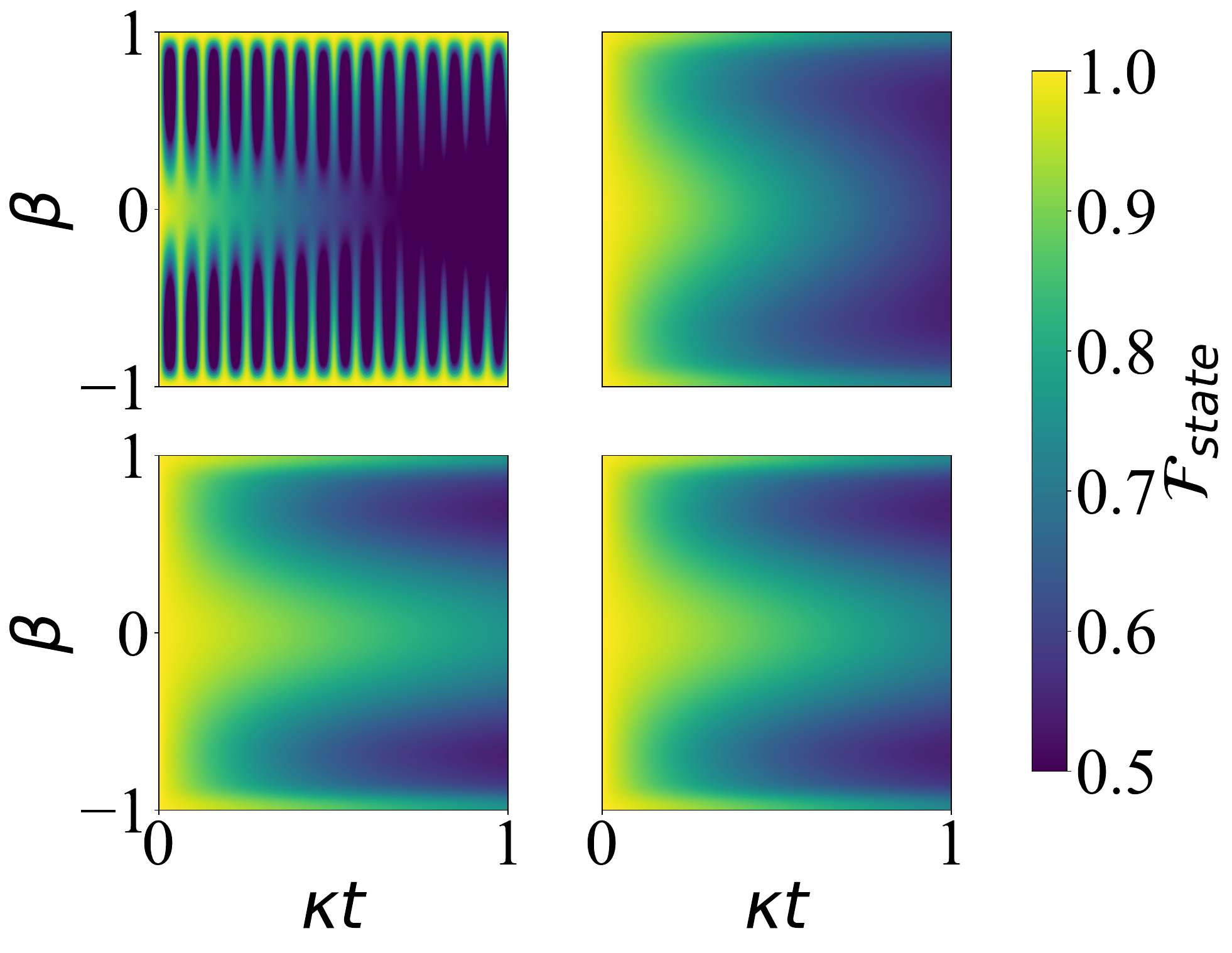}
\caption{\idg{a} State fidelity $\mathcal{F}_{state}$ of the initial state $\ketbra{0}{0}$ as a function of time $t$ and coupling strength $\kappa$ with and without five-qubit QEC. Single and multiple cycle QEC are considered for the analysis. All qubits are coupled to baths with coupling strength $\kappa/\omega=0.01$ and bath temperature $T=0.2$. \idg{b} Heatmap showing the fidelity against time with the same coupling strength and temperature for different initial state $\alpha\ket{0}+\beta\ket{1}$. Four sub-figures (clockwise from top-left) show the results without QEC and for 5-qubit QEC with 1, 5, and 10 cycles, respectively.}
\label{fig3}
\end{figure*}

\begin{figure}[htpb]
\centering
\includegraphics[width=\columnwidth]{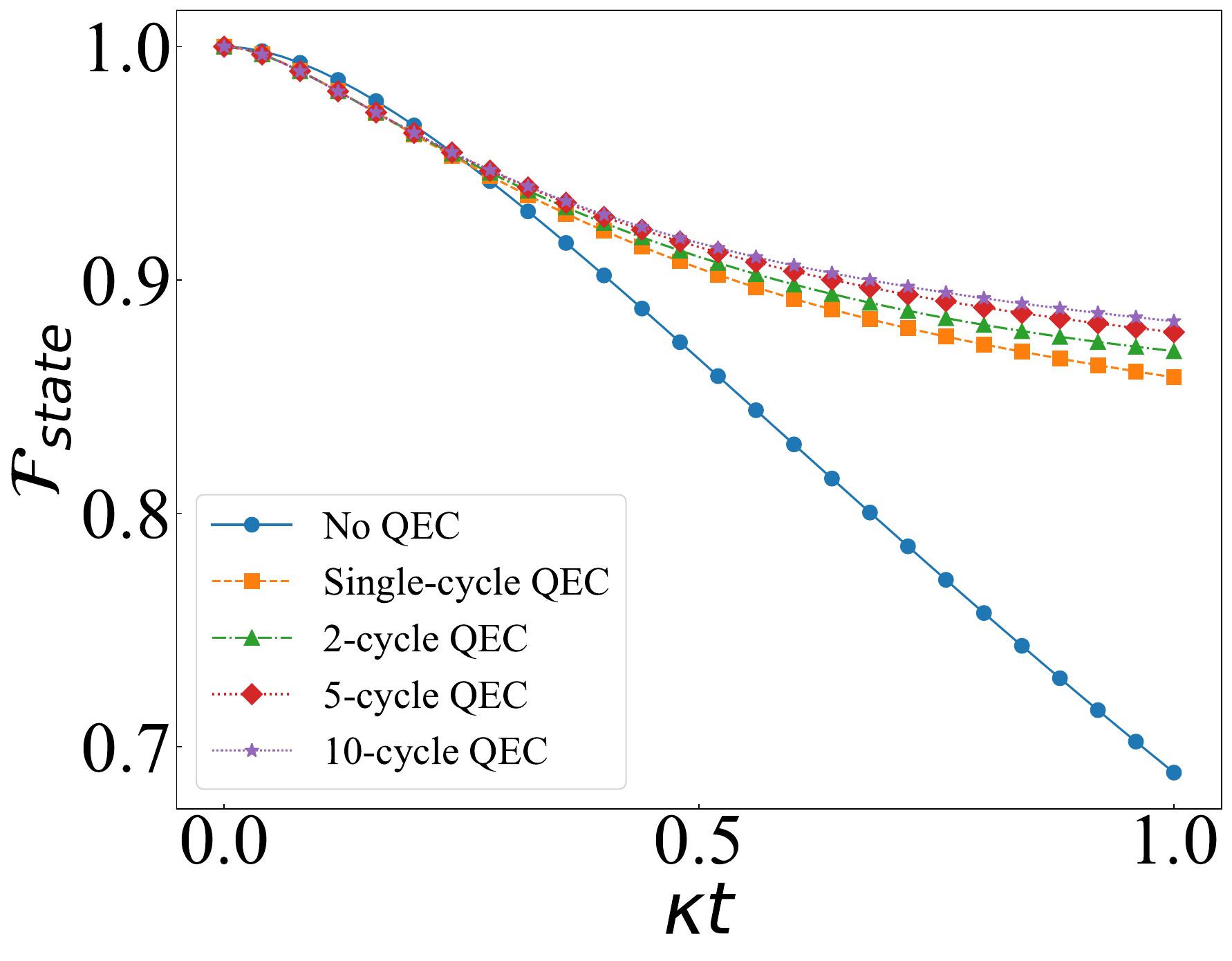}
\hfill
\caption{State fidelity $\mathcal{F}_{state}$ of the initial state $p \ketbra{\psi^-}{\psi^-}+ (1-p) \frac{I}{4}$ for $p=0.5$ as a function of time $t$ and coupling strength $\kappa$ with and without five-qubit QEC. We have considered single as well as multiple cycles QEC. In the weak coupling regime, i.e., $\kappa/\omega=0.01$, all the qubits are considered to be coupled to baths with a temperature $T=0.2$.}
\label{fig4}
\end{figure}

\begin{figure*}[htpb]
\centering
\subfigimg[width=0.49\textwidth]{a)}{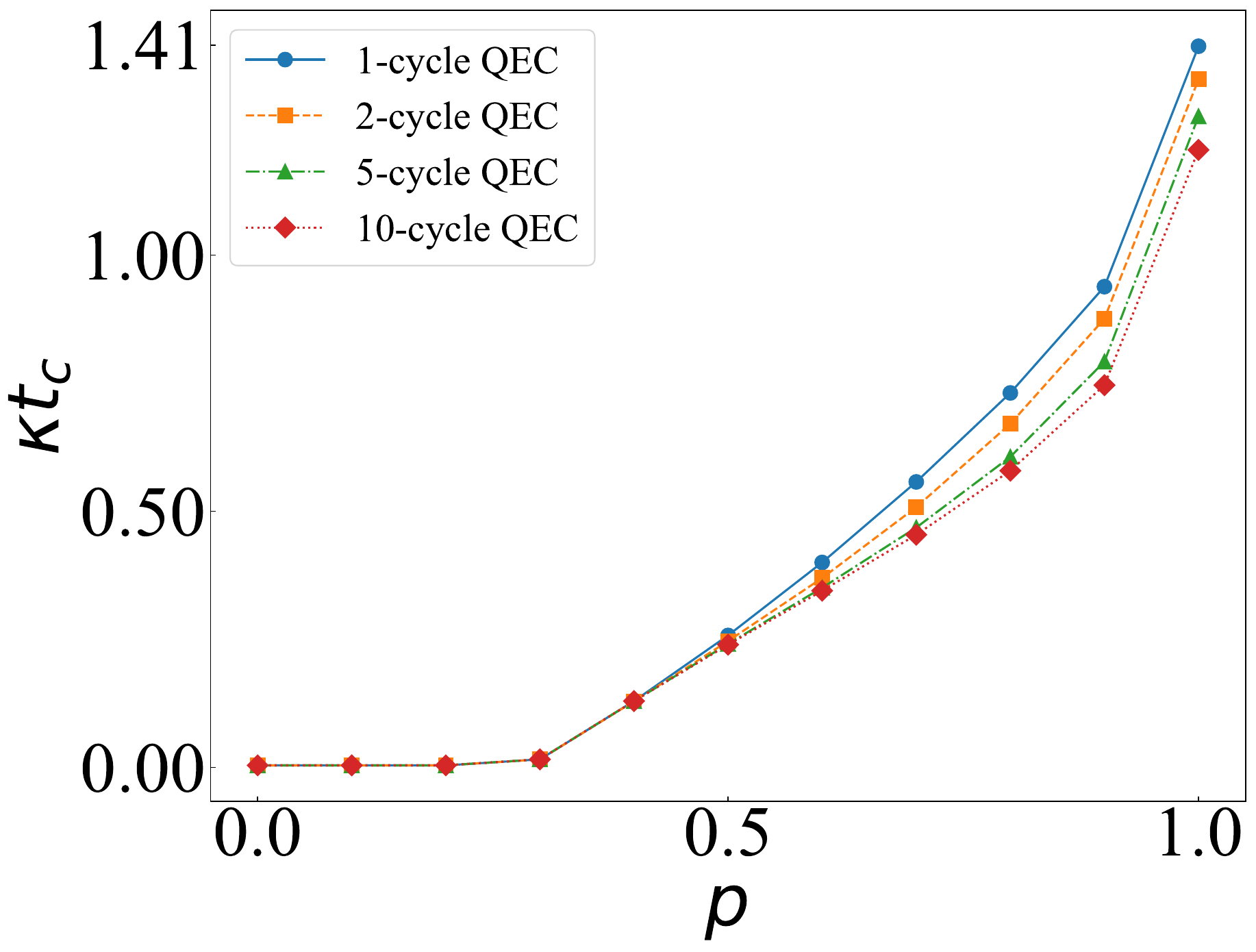}
\hfill
\subfigimg[width=0.49\textwidth]{b)}{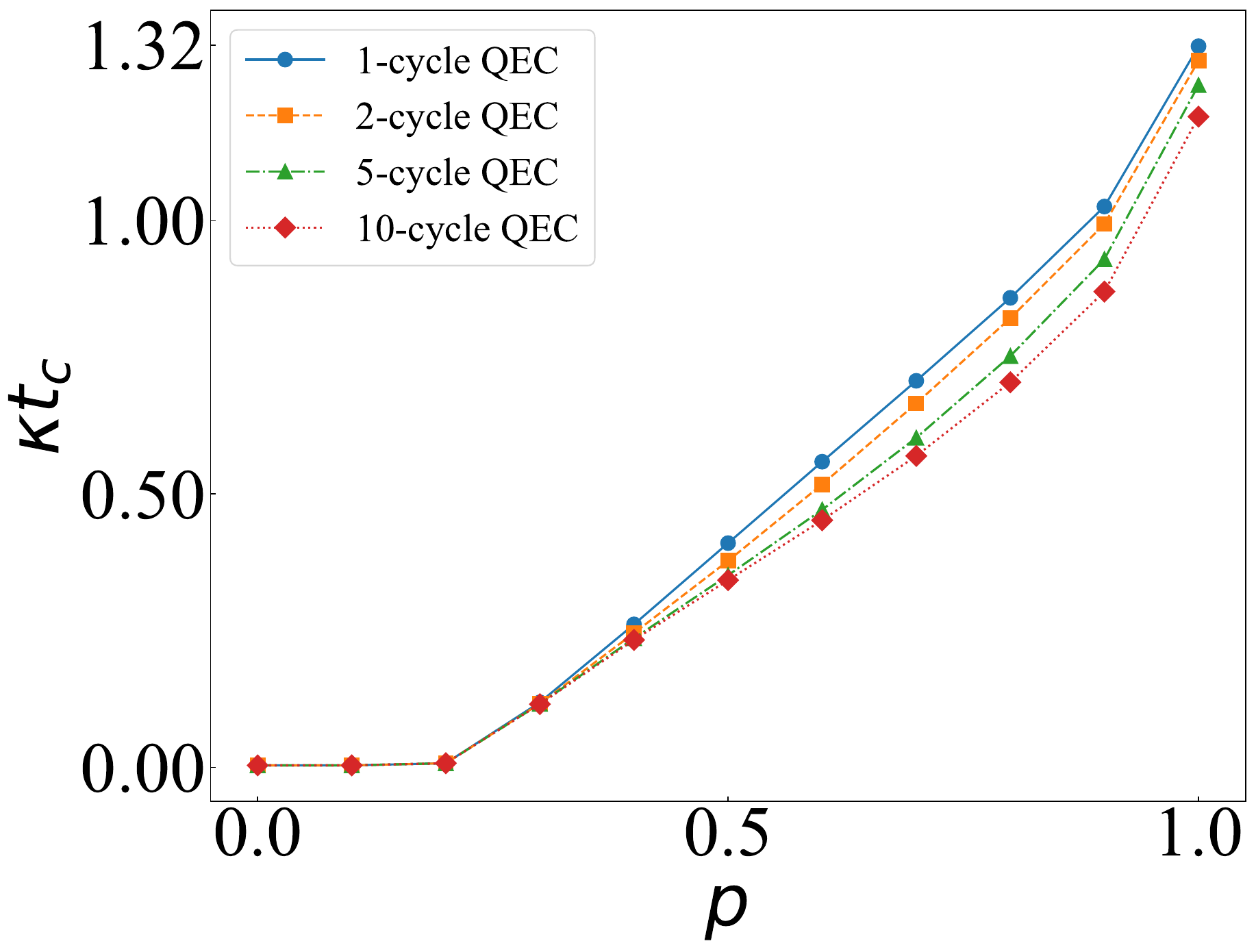}
\caption{The critical values $\kappa t_c$ where QEC fails to improve fidelity are plotted against mixing parameter $p$ of different Werner states $p\ketbra{\psi^-}{\psi^-}+(1-p)\frac{I}{4}$. Single and multiple cycle five-qubit QEC are considered for the process. All qubits are coupled to baths with coupling strength $\kappa/\omega=0.01$, with the bath temperature \idg{a} $T=0.2$ and \idg{b} $T=10$. This indicates that this five-qubit QEC cannot protect entanglements. However, the critical value can be reduced by multiple cycles of QEC.}
\label{fig5}
\end{figure*}

\subsection{Single-logical qubit}

\subsubsection{Low-temperature limit}

In the single-qubit scenario, we employ the five-qubit QEC code to protect the initial quantum state from decoherence induced by its interaction with the environment. The analysis is executed for weak and moderate system–bath coupling. 

For the collective noise case, where both the system qubit and the ancillae are coupled to a common bath ($T$ is identical), we find that in the weak coupling regime ($\kappa/\omega=0.01$), the QEC code significantly enhances the fidelity of the stored quantum state (Fig. \ref{fig3}a with initial state $\ket{0}$). The fidelity exhibits a steady increase with the number of QEC cycles, indicating that successive rounds of error correction effectively suppress decoherence and energy-relaxation effects arising from the system–bath interaction. We encounter a similar trend for a general qubit as an initial pure state, as shown in Fig.~\ref{fig3}b. This demonstrates the stabilizing role of QEC in mitigating noise when the system–environment coupling is sufficiently weak. Even for moderate system–bath coupling ($\kappa/\omega=0.1$), the qualitative behavior of the fidelity remains very similar to the weak-coupling case. Multiple QEC cycles still systematically enhance the fidelity compared with the uncorrected evolution.

In contrast, for local noise, where the system and each ancilla qubit interact with independent baths, the five-qubit code provides no improvement compared to the uncorrected evolution, even in the weak coupling limit. The uncorrelated nature of the local environmental interactions leads to independent error channels that are not efficiently addressed by the collective structure of the five-qubit code. Consequently, the effectiveness of QEC is limited in this regime.
This reflects a structural mismatch: independent noise produces multiple uncorrelated errors that are harder to correct, whereas collective noise induces correlated errors that are more efficiently captured by the code.

\subsubsection{High-temperature limit}

In this regime, as the bath temperature increases, the thermal excitations are enhanced in the environment. 
The temperature dependence of the dynamics arises from the bosonic occupation number 
$n(\omega)=[exp(\omega/k_B T)-1]^{-1}$, which enters the bath correlation functions. For high temperatures, one has $n(\omega)\approx k_B T/\omega$, leading to enhanced excitation and relaxation rates that scale linearly with temperature. This increase in transition rates accelerates decoherence.

For the collective noise configuration, the elevated temperature amplifies correlated fluctuations in the shared bath, resulting in stronger collective dephasing and dissipation processes. In both the weak and moderate coupling cases, the five-qubit code still yields higher fidelities than the uncorrected evolution. We observe a striking feature that the QEC code qualitatively produces the same enhancement as that of the low temperature limit. Thus, this makes the code more robust to the thermal fluctuations, which results in a stronger dephasing.

\subsection{Two logical qubits}

\begin{figure*}[htpb]
\centering
\subfigimg[width=0.49\textwidth]{a)}{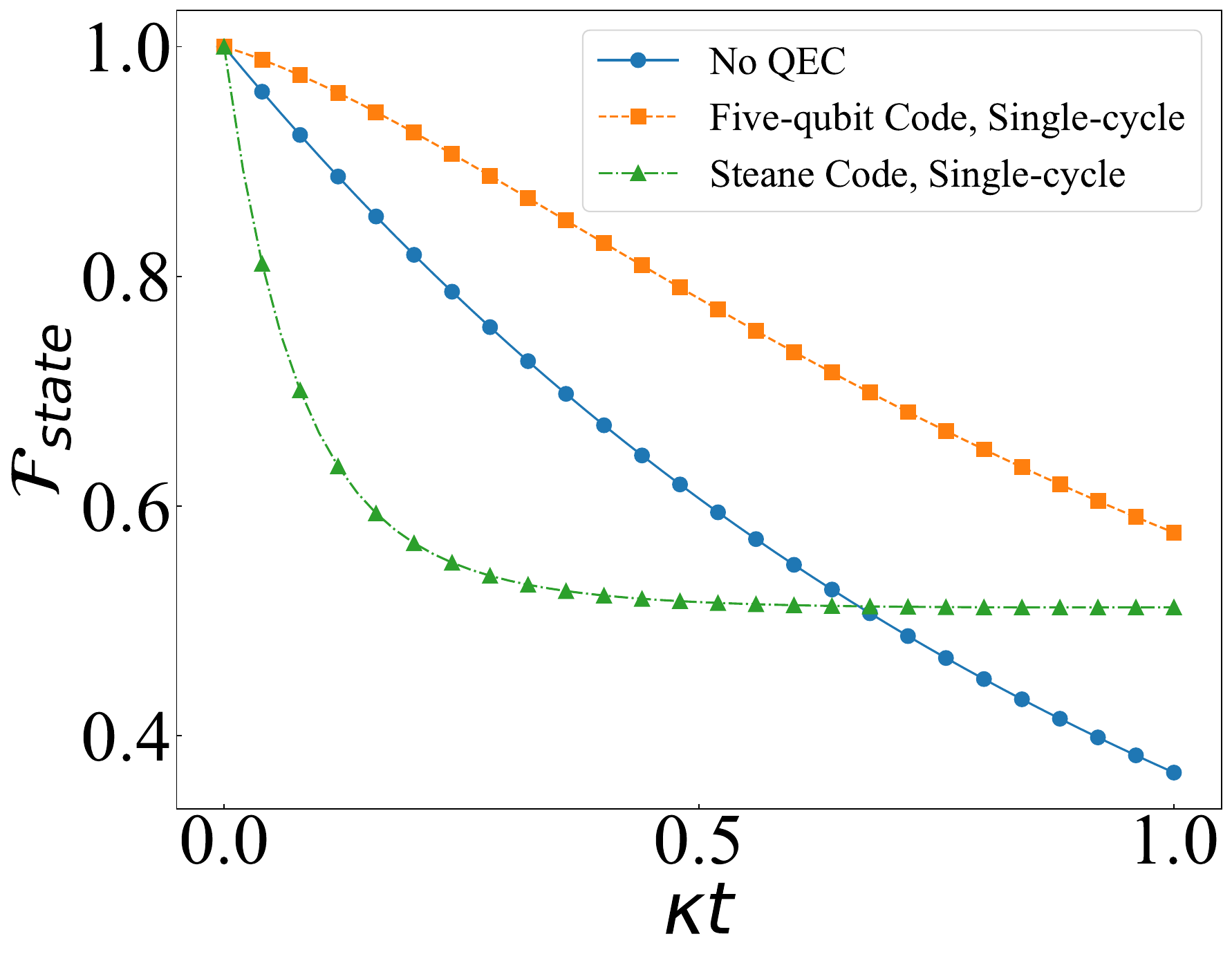}
\hfill
\subfigimg[width=0.49\textwidth]{b)}{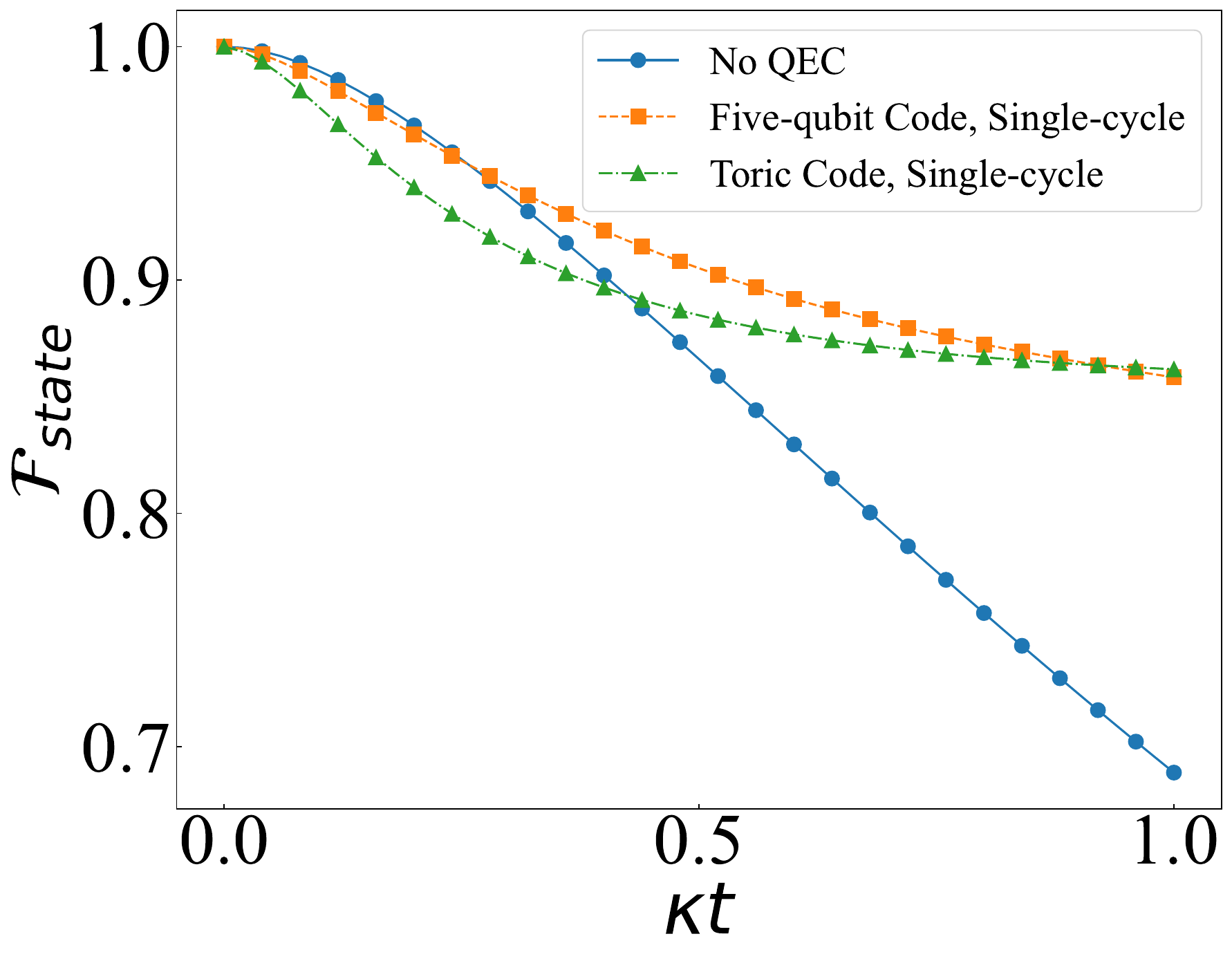}
\caption{\idg{a} State fidelity $\mathcal{F}_{state}$ of the initial state $\ket{0}$ with five-qubit QEC and Steane code. \idg{b} State fidelity $\mathcal{F}_{state}$ of the initial state $p \ketbra{\psi^-}{\psi^-}+ (1-p)\frac{I}{4}$ for $p=0.5$ with five-qubit QEC and Toric code. All qubits are coupled to baths with coupling strength $\kappa/\omega=0.01$, with the bath temperature $T=0.2$.}
\label{fig6}
\end{figure*}

We consider an initial Werner state to study the relation between entanglement and QEC performance. The results reveal a trade-off between QEC advantage and the degree of inter-qubit entanglement (Fig. \ref{fig4}). For weakly entangled or nearly separable states, the application of QEC substantially improves the fidelity compared to the uncorrected dynamics. As the degree of entanglement increases, however, this advantage progressively decreases and creates a small-time window where the QEC codes have less fidelity in state recovery over the uncorrected evolution. %
A crossover point emerges, before which the fidelity of the uncorrected evolution surpasses that of the QEC-protected case. This crossover shifts with increasing entanglement, indicating that highly entangled states couple more strongly to the environment and are therefore more challenging to stabilize using standard QEC codes. These observations highlight the intricate interplay between entanglement, environmental correlations, and the practical limits of error correction in open quantum systems.

To quantify the regime in which QEC remains beneficial for entangled inputs (for high and low temperature limit), in Fig. \ref{fig5}, we plot the critical value $\kappa t_c$ at which the fidelity with five-qubit QEC becomes equal to the uncorrected fidelity, as a function of the mixing parameter $p$ of the initial Werner state. For both the high- and low-temperature limit and the number of QEC cycles, $\kappa t_c$ sets the boundary beyond which the error-corrected evolution provides a net advantage over the bare dynamics. The curves show a systematic dependence on the degree of entanglement: as $p$ increases and the state approaches the maximally entangled Bell state, the window in $\kappa t$ over which QEC improves the state shrinks, indicating that highly entangled states are more fragile to the combined action of system–bath coupling and thermal noise. Increasing the number of QEC cycles shifts $\kappa t_c$ so that the beneficial region extends over a broader range of evolution times for all $p$, but it does not eliminate the overall trend that strongly entangled states are harder to protect than weakly entangled or nearly separable states using traditional QEC codes. We restrict Fig. \ref{fig5} to the weak-coupling case $\kappa/\omega =0.01 $, as the larger system-bath coupling, $\kappa/\omega =0.1 $, displays very similar behavior and would not change the qualitative conclusions.

The emergence of a critical evolution time can be understood as a crossover between two competing effects: at short times, the error accumulated due to system–bath interaction remains small, and the application of QEC introduces an effective overhead due to encoding and recovery operations. At longer times, however, the accumulated error grows approximately linearly with time, and beyond a characteristic scale $t_c \propto 1/\Gamma_{\rm eff}$ (see Appendix~\ref{AppendixG}), QEC begins to provide a net advantage. The dependence of this critical time on the Werner-state parameter reflects the increased fragility of highly entangled states to environmental decoherence.

\subsection{Comparative analysis}

The comparative analysis of different QEC codes for the weak and moderate system–bath coupling reveals that the five-qubit code provides a markedly superior performance in preserving the quantum state compared to both the Steane and toric codes. We emphasize that this comparison is carried out under identical microscopic noise conditions and is not intended as a resource-matched benchmark.  The performance comparison of the five-qubit code, with the Steane and Toric code, is presented in Fig.~\ref{fig6} (see Appendix~\ref{appendix:F} with initial state $\ket{+}$). For this analysis, we take initial states as $\ket{0}$ for a single logical qubit when comparing the five-qubit code with the Steane code, and the Werner state for two logical qubits when comparing with the toric code. For both the coupling strength and for high and low temperature limits, the five-qubit code shows a significant advantage over other QEC codes. This advantage arises from its minimal code size and fully entangled structure, which enables faster recovery and reduced accumulation of uncorrected errors.

In contrast, the Steane and toric codes, while effective in fault-tolerant and topological architectures, respectively, exhibit slower recovery dynamics and greater sensitivity to environmental perturbations in the same regime. Their larger error spaces and higher overhead make them less efficient in combating weak but frequent noise processes.

\section{Conclusion}\label{Conc}

To summarize, we have presented an investigation of the \textit{resilience} of quantum information in OQSs subjected to realistic environmental interactions. By explicitly modeling the system–bath coupling within a microscopic Hamiltonian framework and analyzing the resulting dynamics through the master equation formalism, we have quantitatively assessed the performance of several leading QEC codes, namely, the five-qubit, Steane, and toric codes, for different coupling strengths and bath temperatures. This approach goes beyond conventional channel-based analyses by capturing the interplay between decoherence, dissipation, and non-Markovian effects inherent to realistic quantum environments.

Compared to standard phenomenological noise channels, the present approach derives decoherence from an explicit system–bath Hamiltonian, enabling a direct connection between physical parameters such as temperature and coupling strength and the resulting noise processes. This allows us to capture temperature-dependent effects, correlated noise structures, and time-dependent dynamics within a unified framework. While in the weak-coupling regime, the resulting dynamics may approximate familiar channel models, the microscopic formulation provides additional physical insight into how environmental interactions shape the performance of quantum error correction.

Our results show that the five-qubit code provides a significant \textit{enhancement} in preserving quantum state fidelity for both weak and moderately strong system–bath coupling in the low and high temperature regimes. It is better suited compared to the Steane code and small toric code for near-term quantum devices operating in open-system conditions, where resources are constrained and noise correlations are significant. However, the toric codes with larger lattice sizes may increase the performance. Multiple QEC cycles further suppress decoherence, demonstrating that repetitive correction effectively counteracts the cumulative effects of environmental noise within this coupling range. For two-qubit entangled states, a crossover behavior emerges, where QEC protection becomes less efficient as the degree of entanglement increases, indicating that stronger inter-qubit correlations can enhance the susceptibility to environmental decoherence. Moreover, our analysis reveals the limitations of conventional QEC codes when faced with high thermal noise and larger system–bath coupling, highlighting the need for modified or adaptive correction strategies that account for the underlying physical noise structure.

Our framework provides a bridge between microscopic open-system dynamics and QEC performance at the logical level. While it does not explicitly construct logical noise channels or address asymptotic scalability, it offers a physically grounded approach to assessing how different code architectures respond to realistic system–bath interactions. Extensions incorporating explicit logical channel reconstruction, renormalization effects, and large-scale code limits would further deepen the connection between microscopic noise modeling and fault-tolerant quantum computation.

Building on these results, several promising research directions emerge. First, it would be valuable to extend this framework to the larger toric codes, \textit{surface codes} and \textit{dynamical QEC schemes} that adapt correction intervals or code parameters based on the evolving decoherence rates inferred from the environment. Second, investigating \textit{hybrid QEC strategies} that integrate bosonic codes with discrete stabilizer codes may offer improved protection in regimes dominated by thermal and dissipative noise. Finally, exploring the thermodynamic cost and entropy production associated with QEC operations in open systems could establish a deeper connection between information recovery, dissipation, and the fundamental limits of reversibility in quantum dynamics. Such extensions would not only advance the theoretical understanding of QEC in realistic environments but also provide \textit{practical guidelines for implementing noise-resilient quantum computation and communication protocols} in the next generation of quantum devices.

Alternative approaches based on bosonic or continuous-variable encodings, particularly those employing quantum non-Gaussian (QNG) states, have recently attracted significant attention. Such encodings exploit larger Hilbert spaces and phase-space structure to provide enhanced resilience against dissipative and thermal noise processes. Recent studies~\cite{adhikary2025coherence,58jrl1x6,fong2025engineering} demonstrate that QNG resources can play a crucial role in identifying and mitigating noise-induced degradation in open quantum systems. Exploring the integration of microscopic noise modeling with such encoding strategies represents an important direction for future work.

\section{Acknowledgement}
NB and PC would like to thank Saikat Sur of the Indian Institute of Mathematical Science, Chennai, and Avijit Misra of the Indian Institute of Technology, Dhanbad, for their valuable inputs and suggestions. P.C. acknowledges the support from the International Postdoctoral Fellowship from the Ben May Center for Theory and Computation.

\bibliography{apssamp.bib}

\onecolumngrid
\appendix
\newpage

\setcounter{equation}{0}
\renewcommand{\theequation}{A\arabic{equation}}
\section{Bath Response functions \label{appendix:A}}

Using the standard definition of the bath correlation functions, which encapsulate the response of the thermal reservoir to the system-bath coupling, we write: $$\Phi_{\alpha \alpha'}(\tau) = Tr_B(B_\alpha(\tau)B_{\alpha'}(0)\rho_B),$$ where $\rho_B$ denotes the thermal state of the bath, and $B_{\alpha,j}$ represents the interaction-picture time evolution of the bath operators under the free bath Hamiltonian $H_B$. Using this definition and expanding the operator $B_{\alpha,j}$ in terms of the bosonic creation and annihilation operators, the two-point bath correlation function can be explicitly evaluated as:

\begin{align}\nonumber
\Phi_{\alpha\alpha'}(\tau)&=\langle B_{\alpha,j}(t-\tau)B_{\alpha',j}(t)\rangle_B\\\nonumber
&=\tr\{B_{\alpha,j}(t-\tau)B_{\alpha',j}(t)\rho_B\}\\
&=\tr\left(e^{iH_B(t-\tau)}B_{\alpha,j}e^{-iH_B(t-\tau)}e^{iH_Bt}B_{\alpha',j}e^{-iH_Bt}\rho_B\right)\\\nonumber
&=\tr\left(e^{-iH_B\tau}B_{\alpha,j}e^{iH_B\tau}B_{\alpha',j}\rho_B\right)\\\nonumber
&=\sum_{kk'}\left(g_{k,j}g_{k',j}\Gamma(\tau;k,k')\right)
\end{align}
where the kernel $\Gamma(\tau;k,k')$ encodes the dynamical contributions of the bath modes and depends on the indices $\alpha$, $\alpha'$ as follows:
\begin{align}
\Gamma(\tau;k,k')&=\begin{cases}
e^{i\tau\Omega_{k,j}}\langle b_{k,j}b_{k',j}\rangle_B&\alpha=\alpha'=1\\
e^{i\tau\Omega_{k,j}}\langle b_{k,j}b_{k',j}^\dagger\rangle_B&\alpha=1,\alpha'=2\\
e^{-i\tau\Omega_{k,j}}\langle b_{k,j}^\dagger b_{k',j}\rangle_B&\alpha=2,\alpha'=1\\
e^{-i\tau\Omega_{k,j}}\langle b_{k,j}^\dagger b_{k',j}^\dagger\rangle_B&\alpha=\alpha'=2
\end{cases}\\
&=\begin{cases}
0&\alpha=\alpha'\\
\delta_{kk'}e^{i\Omega_{k,j}\tau}\left(n_j(\omega_j)+1\right)&\alpha=1,\alpha'=2\\
\delta_{kk'}e^{-i\Omega_{k,j}\tau}\left(n_j(\omega_j)\right)&\alpha=2,\alpha'=1
\end{cases}
\end{align}
where $n_j(\Omega)=\frac{1}{e^{\frac{\Omega}{k_BT_j}}-1}, k_B$ is Boltzmann constant, $T_j$ is the temperature of $j$-th bath. Therefore, substituting these thermal averages yields the simplified form:
\begin{equation}
\Phi_{\alpha\alpha'}(\tau)=\begin{cases}
0,&\alpha=\alpha'\\
\sum_k|g_{k,j}|^2e^{i\Omega_{k,j}\tau}\left(n_j(\omega_j)+1\right),&\alpha=1,\alpha'=2\\
\sum_k|g_{k,j}|^2e^{-i\Omega_{k,j}\tau}\left(n_j(\omega_j)\right),&\alpha=2,\alpha'=1
\end{cases}
\end{equation}

with $\sum_{k} \vert g_{k,j} \vert^2$, is a finite positive number dependent on the nature of interaction between the bath and the system qubits. These correlation functions serve as the fundamental inputs for characterizing both dissipative and decoherence dynamics in open quantum systems.

\setcounter{equation}{0}
\renewcommand{\theequation}{B\arabic{equation}}
\section{Solution to the Master equation \label{appendix:B}}

The time-nonlocal master equation, commonly referred to as the Nakajima-Zwanzig master equation~\cite{breuer2007theory}, governs the evolution of the reduced density operator $\rho_S (t)$ of the system. To second order in the system-bath coupling strength, this equation encapsulates the non-Markovian dynamics arising from the system's interaction with its environment, and is expressed as:

\begin{eqnarray}
  \dot{\rho_S}(t) &&= -i[H_S(t), \rho_S(t)]+ \int^t_{s = 0} ds~ \text{tr}_B \{ [ H_{SB}(t) \rho_S(t)\rho_B,  H_{SB}(t-s)] \} + h.c. \nonumber\\
  && = -i[H_S(t), \rho_S(t)]+ \int^t_{s = 0} ds~ \text{tr}_B \{ H_{SB}(t) \rho_S(t)\rho_B  H_{SB}(t-s) -  H_{SB}(t-s)  H_{SB}(t) \rho_S(t)\rho_B \} + h.c. \nonumber\\
  && = -i[H_S(t), \rho_S(t)]+  \Bigg[ S_1(t) \rho_S(t) \Big(\int^t_0 ds~  S_2(t-s) \Phi_{21}(-s)\Big) +  S_2(t) \rho_S(t) \Big(\int^t_0 ds~ S_1(t-s) \Phi_{12}(-s)\Big)\nonumber\\
  && - \Big( \int^t_0 ds~\Phi_{12}(-s) S_1(t-s)\Big)  S_2(t) \rho_S(t) - \Big( \int^t_0 ds~\Phi_{21}(-s)  S_2(t-s)\Big)  S_1(t) \rho_S(t)\Bigg] + h.c. \nonumber\\
  && = -i[H_S(t), \rho_S(t)]+ \Bigg[ S_1(t) \rho_S(t) W_2(t) +  S_2(t) \rho_S(t) W_1(t) - W_1(t)  S_2(t) \rho_S(t) - W_2(t)  S_1(t) \rho_S(t) \Bigg]  + h.c. \nonumber\\
 \end{eqnarray}

where the operators $ W_1(t)$ and $ W_2(t)$ encapsulate the bath-induced memory effects and are defined as:
\begin{eqnarray}
 &&W_1(t) = \int^t_0 ds ~S_1(t-s) \Phi_{12}(-s),\nonumber\\
&&W_2(t) = \int^t_0 ds ~S_2(t-s) \Phi_{21}(-s),\nonumber\\
\end{eqnarray}

where, $S(t)$ are given in Eq.~\eqref{eqnSS}. The bath correlation functions $\phi_{ij}(-s)$ encapsulate the statistical properties of the environment and mediate the non-Markovian features of the dynamics.

\subsection*{Master equation for two-qubit logical system QEC}
The $S_{1(2),j}$ operator as defined in Eq.~\eqref{eqnSS} for this case becomes

\begin{align}\nonumber
S_{1(2),j}(t)&=e^{iH_St}S_{1(2),j}e^{-iH_St}\\\nonumber
&=e^{-it\frac{\omega_j}{2}\sigma_{zj}}\sigma_j^{+(-)}e^{it\frac{\omega_j}{2}\sigma_{zj}}\\
&=\left[\cos{t\omega_j}-(+)i\sin{t\omega_j}\right]\sigma_j^{+(-)}
\end{align}

Now
\begin{align}
&\int_0^t d\tau\tr_B\left[H^I_j(t)\rho_S(t)\otimes\rho_B,H^I_j(t-\tau)\right]\notag\\ \nonumber
=&\int_0^td\tau\sum_{\alpha\alpha'}\left[S_{\alpha,j}(t)\rho_S(t), S_{\alpha',j}(t-\tau)\right]\langle B_{\alpha',j}(t-\tau)B_{\alpha,j}(t)\rho_B\rangle_B\\ \nonumber
=&\sum_k|g_{k,j}|^2\Big((\sigma^+_j\rho_S(t)\sigma_j^--\sigma^-_j\sigma_j^+\rho_S(t))\frac{in_j(\omega_j)}{\omega_j+\Omega_{k,j}}(e^{-it(\omega_j+\Omega_{k,j})}-1)\\
&+(\sigma^-_j\rho_S(t)\sigma_j^+-\sigma^+_j\sigma_j^-\rho_S(t))\frac{-i(n_j(\omega_j)+1)}{\omega_j+\Omega_{k,j}}(e^{it(\omega_j+\Omega_{k,j})}-1)\Big)
\end{align}

\setcounter{equation}{0}
\renewcommand{\theequation}{C\arabic{equation}}
\section{Encoding and decoding of five-qubit code \label{appendix:C}}

In five-qubit code~\cite{laflamme1996perfect}, the encoding operation $U$ encodes one physical qubit into the logical space spanned by $\{\ket{0_L},\ket{1_L}\}$ using five physical qubits, where
\begin{equation}
\begin{aligned}
\ket{0_L}&=U\ket{00000}=\frac{1}{2\sqrt{2}}(-\ket{00000}+\ket{00110}+\ket{01001}+\ket{01111}-\ket{10011}+\ket{10101}+\ket{11010}+\ket{11100})\\
\ket{1_L}&=U\ket{00100}=\frac{1}{2\sqrt{2}}(-\ket{11111}+\ket{11001}+\ket{10110}+\ket{10000}+\ket{01100}-\ket{01010}-\ket{00101}-\ket{00011}).
\end{aligned}
\end{equation}
Therefore, for any single qubit density matrix $\rho$, we can write the encoding as
\begin{equation}
\rho_L=\mathcal{E}(\rho)=U\kb{00}{00}\otimes\rho\otimes\kb{00}{00} U^\dagger.
\end{equation}
The state recovery stage consists of decoding, error detection, and correction. The decoding operation is the reverse of the encoding operation. That is, after decoding, we get,
\begin{equation}
\hat\rho=\mathcal{D}(\hat\rho_L)=U^\dagger\hat\rho_LU,
\end{equation}
where $\hat\rho_L$ is the erroneous state.

The error detection consists of measuring the auxiliary qubits $j=1, 2, 4, 5$. This measurement is called \textit{syndrome measurement}. Depending on these measurement results, called \textit{error syndrome}, Pauli operations are applied to the main qubit ($j=3$) to correct the error. We can combine these detections and corrections and write the operation in terms of the Kraus map as
\begin{equation}
\kb{00}{00}\otimes\rho'\otimes\kb{00}{00}=\mathcal{R}(\hat\rho)=\sum_{k=0}^{15}R_k\hat\rho R_k^\dagger,
\end{equation}
where the operators $\{R_k\}$ are given by Ref.~\cite{babu2023quantum} as,
\begin{align*}
R_0&=\kb{00}{00}\otimes\sigma_0\otimes\kb{00}{00},&R_1&=\kb{00}{00}\otimes\sigma_z\otimes\kb{00}{01},\\
R_2&=\kb{00}{00}\otimes\sigma_0\otimes\kb{00}{10},&R_3&=\kb{00}{00}\otimes\sigma_0\otimes\kb{00}{11},\\
R_4&=\kb{00}{01}\otimes\sigma_0\otimes\kb{00}{00},&R_5&=\kb{00}{01}\otimes\sigma_z\otimes\kb{00}{01},\\
R_6&=\kb{00}{01}\otimes\sigma_x\otimes\kb{00}{10},&R_7&=\kb{00}{01}\otimes\sigma_x\otimes\kb{00}{11},\\
R_8&=\kb{00}{10}\otimes\sigma_0\otimes\kb{00}{00},&R_9&=\kb{00}{10}\otimes\sigma_x\otimes\kb{00}{01},\\
R_{10}&=\kb{00}{10}\otimes\sigma_z\otimes\kb{00}{10},&R_{11}&=\kb{00}{10}\otimes\sigma_x\otimes\kb{00}{11},\\
R_{12}&=\kb{00}{11}\otimes\sigma_z\otimes\kb{00}{00},&R_{13}&=\kb{00}{11}\otimes\sigma_x\sigma_z\otimes\kb{00}{01},\\
R_{14}&=\kb{00}{11}\otimes\sigma_x\otimes\kb{00}{10},&R_{15}&=\kb{00}{11}\otimes\sigma_z\otimes\kb{00}{11},\\
\end{align*}
where $\sigma_0$ is the identity operator and $\sigma_x,\sigma_z$ are Pauli operators.

\setcounter{equation}{0}
\renewcommand{\theequation}{D\arabic{equation}}
\section{Encoding and decoding of CSS code: Steane Code \label{appendix:D}}

\begin{table}[h]
\centering
\begin{tabular}{c|c|c}
\hline
\textbf{Syndrome ($G_1G_2G_3G_4G_5G_6$)}&\textbf{Type of Error}&\textbf{Correction Operation}\\
\hline
000000&No Error&$IIIIIII$\\
\hline
000001&$B1$&$XIIIIII$\\
000010&$B2$&$IXIIIII$\\
000011&$B3$&$IIXIIII$\\
000100&$B4$&$IIIXIII$\\
000101&$B5$&$IIIIXII$\\
000110&$B6$&$IIIIIXI$\\
000111&$B7$&$IIIIIIX$\\
\hline
001000&$P1$&$ZIIIIII$\\
010000&$P2$&$IZIIIII$\\
011000&$P3$&$IIZIIII$\\
100000&$P4$&$IIIZIII$\\
101000&$P5$&$IIIIZII$\\
110000&$P6$&$IIIIIZI$\\
111000&$P7$&$IIIIIIZ$\\
\end{tabular}
\caption{Error correction using 7-qubit Steane code. A single qubit error can be detected and corrected from syndrome measurement. The measurement results and the corresponding correction operations are given here. $BN$ and $PN$ denotes bit-flip and phase-flip errors on $N$-th qubit.}
\label{tab:Steane_corr}
\end{table}

Steane code~\cite{steane1996multiple,steane1996error} is a seven-qubit stabilizer code that uses the stabilizer formalism. A quantum error correcting code (QECC) with a stabilizer group, which stabilizes the logical codewords, is called a stabilizer code. Each stabilizer code has a set of stabilizer generators that can generate the whole stabilizer group. From this information, one can easily derive the logical codewords and logical operations. The error syndromes are given by the measurement of stabilizer generators. The six stabilizer generators of the Steane code are $\{G_1=IIIXXXX, G_2=IXXIIXX, G_3=XIXIXIX, G_4=IIIZZZZ, G_5=IZZIIZZ, G_6=ZIZIZIZ\}$ and the logical Pauli-$X$ and Pauli-$Z$ operations are $XXXXXXX$ and $ZZZZZZZ$ respectively. Thus the computational basis for the logical space would be $\{\ket{0_L},\ket{1_L}\}$ where,
\begin{equation}
\begin{aligned}
\ket{0_L}&=\frac{1}{2\sqrt{2}}(\ket{0000000}+\ket{1010101}+\ket{0110011}+\ket{1100110}+\ket{0001111}+\ket{1011010}+\ket{0111100}+\ket{1101001})\\
\ket{1_L}&=\frac{1}{2\sqrt{2}}(\ket{1111111}+\ket{0101010}+\ket{1001100}+\ket{0011001}+\ket{1110000}+\ket{0100101}+\ket{1000011}+\ket{0010110})
\end{aligned}
\end{equation}

A single-qubit error can be easily detected and corrected using the Steane code. Table~\ref{tab:Steane_corr} shows the syndrome measurement results and the corresponding error type along with the correction operations.

\setcounter{equation}{0}
\renewcommand{\theequation}{E\arabic{equation}}
\section{Encoding and decoding of Toric code \label{appendix:E}}

\begin{figure}[h]
\centering
\includegraphics[width = 0.5\textwidth]{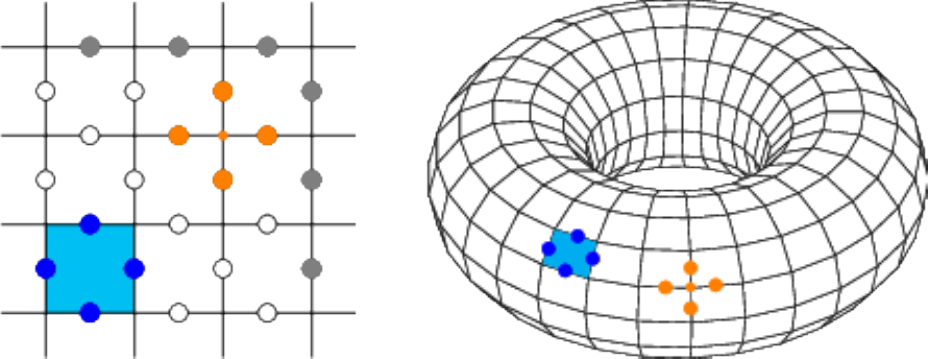}
\caption{Lattice representation of Toric code ($L=3$) and its embedding into a three-dimensional torus.  Circles denote qubits, with gray circles as repeated qubits on boundaries. Vertices and plaquettes are duals of each other, and they denote stabilizers of the code. Two of such stabilizers are shown here in cyan (plaquette) and orange (vertex). One stabilizer is given by applying Pauli-$Z$ on the qubits (shown in blue) adjacent to the plaquette, and the other is given by applying Pauli-$X$ on the qubits (shown in orange) adjacent to the vertex.}
\label{fig:Toric}
\end{figure}

\begin{figure}
\centering
\includegraphics[width = 0.5\textwidth]{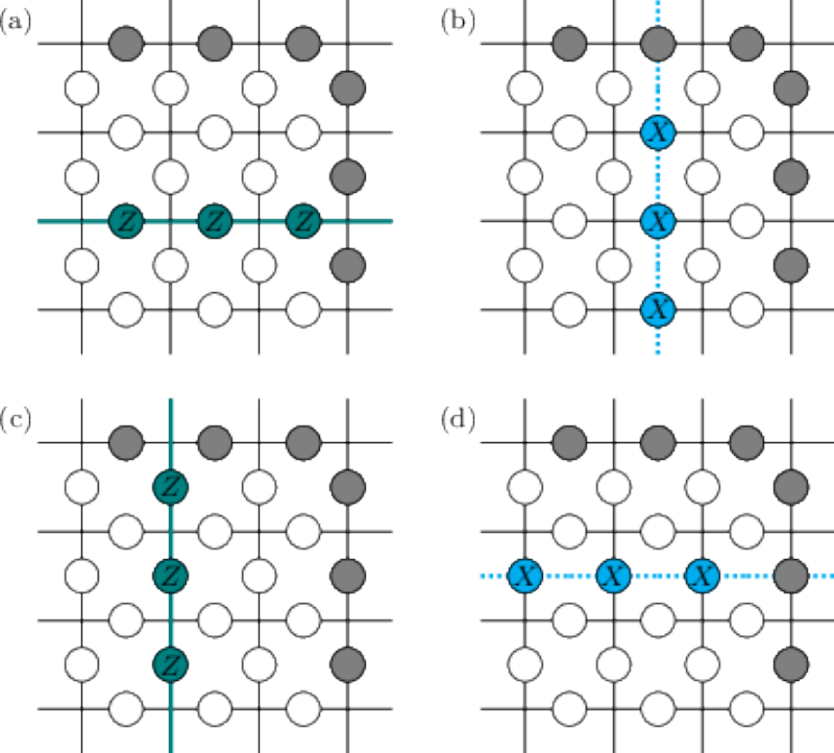}
\caption{Four non-trivial independent loops in a $3\times3$ toric code representing four logical Pauli operators: (a) $Z_1$, Logical $Z$ operator for qubit 1; (b) $X_1$, Logical $X$ operator for qubit 2; (c) $Z_2$, Logical $Z$ operator for qubit 1; (d) $X_2$, Logical $X$ operator for qubit 2.}
\label{fig:Toric_loop}
\end{figure}

The toric code~\cite{kitaev2003fault} is a $[[2L^2,2, L]]$ QEC code, corresponding to an $L\times L$ lattice. Here, 2 logical qubits are encoded using $2L^2$ physical qubits with code distance $L$. Therefore this code can detect any $L-1$-qubit error and correct up to $\lfloor\frac{L-1}{2}\rfloor$-qubit error. A lattice and a three-dimensional torus representation are given in Fig.~\ref{fig:Toric} for $L=3$. This is a stabilizer code. The stabilizer generators are given by applying Pauli-$Z$ on each qubit adjacent to the plaquettes and Pauli-$X$ on each qubit adjacent to the duals of the plaquettes. As the vertices are the duals of the plaquettes, Pauli-$X$ can be applied on the qubits adjacent to the vertices to get stabilizer generators. These are the plaquette and the vertex (star) operators, respectively. In Fig.~\ref{fig:Toric}, one plaquette operator and one vertex operator are shown in blue and red, respectively. There are $L^2-1$ generators of each type, totaling $2(L^2-1)$ generators. The logical Pauli operators are given by non-trivial loops. There are a total of 5 independent loops in a torus. Out of that one is trivial, which can be deformed into a point without breaking it, giving the identity operator. The other four independent loops shown in Fig.~\ref{fig:Toric_loop} give four independent logical Pauli operators.

\begin{table}[h]
\centering
\begin{tabular}{c|c|c}
\hline
\textbf{Syndrome ($G_1G_2G_3G_4G_5G_6$)}&\textbf{Type of Error}&\textbf{Correction Operation}\\
\hline
000000&No Error&$IIIIIII$\\
\hline
000001&$B7,B8$&$IIIIIIXI,IIIIIIIX$\\
000010&$B2,B6$&$IXIIIIII,IIIIIXII$\\
000101&$B1,B5$&$XIIIIIII,IIIIXIII$\\
000110&$B3,B4$&$IIXIIIII,IIIXIIII$\\
\hline
001000&$P5,P6$&$IIIIZIII,IIIIIZII$\\
010000&$P4,P8$&$IIIZIIII,IIIIIIIZ$\\
101000&$P3,P7$&$IIZIIIII,IIIIIIZI$\\
110000&$P1,P2$&$ZIIIIIII,IZIIIIII$\\
\end{tabular}
\caption{Error correction using $[[8,2,2]]$ Toric code. The measurement results and the corresponding correction operations are given here. $BN$ and $PN$ denotes bit-flip and phase-flip errors on $N$-th qubit.}
\label{tab:Toric_corr}
\end{table}

For $L=2$, $2(2^2-1)=6$ generators are $\{G_1=XXXIIIXI,G_2=XXIXIIIX,G_3=IIXIXXXI,G_4=ZIZZZIII,G_5=IZZZIZII,G_6=ZIIIZIZZ\}$. The logical Pauli operators are given by $\{Z_1=ZZIIIIII,X_1=XIIIXIII,Z_2=IIZIIIZI,X_2=IIXXIIII\}$. The computational basis for the code is given by $\{\ket{00}_L,\ket{01}_L,\ket{10}_L,\ket{11}_L\}$ where,
\begin{equation}
\allowdisplaybreaks
\begin{aligned}
\ket{00}_L&=\frac{1}{2\sqrt2}\left[\ket{0}+\ket{29}+\ket{46}+\ket{51}+\ket{204}+\ket{209}+\ket{226}+\ket{255}\right],\\
\ket{01}_L&=\frac{1}{2\sqrt2}\left[\ket{68}+\ket{89}+\ket{106}+\ket{119}+\ket{136}+\ket{149}+\ket{166}+\ket{187}\right],\\
\ket{10}_L&=\frac{1}{2\sqrt2}\left[\ket{3}+\ket{30}+\ket{45}+\ket{48}+\ket{207}+\ket{210}+\ket{225}+\ket{252}\right],\\
\ket{11}_L&=\frac{1}{2\sqrt2}\left[\ket{71}+\ket{90}+\ket{105}+\ket{116}+\ket{139}+\ket{150}+\ket{165}+\ket{184}\right]
\end{aligned}
\end{equation}

The syndrome measurement results and the corresponding correction operations are given in Table~\ref{tab:Toric_corr}. From this table, it can be seen that for each error syndrome, there are two possible errors and two possible correction operations. Therefore, it is not possible to correct the error completely. One possible way may be the random application of the correction operators with a probability of 0.5. However, it being a probabilistic approach, it cannot ensure error correction. By taking $L\geq3$, $\lfloor\frac{L-1}{2}\rfloor\geq\lfloor\frac{3-1}{2}\rfloor=1$-qubit error correction can be ensured.

\setcounter{equation}{0}
\renewcommand{\theequation}{F\arabic{equation}}

\section{Comparative analysis of the different QEC codes \label{appendix:F}}

\begin{figure*}[htpb]
\centering
\subfigimg[width=0.49\textwidth]{a)}{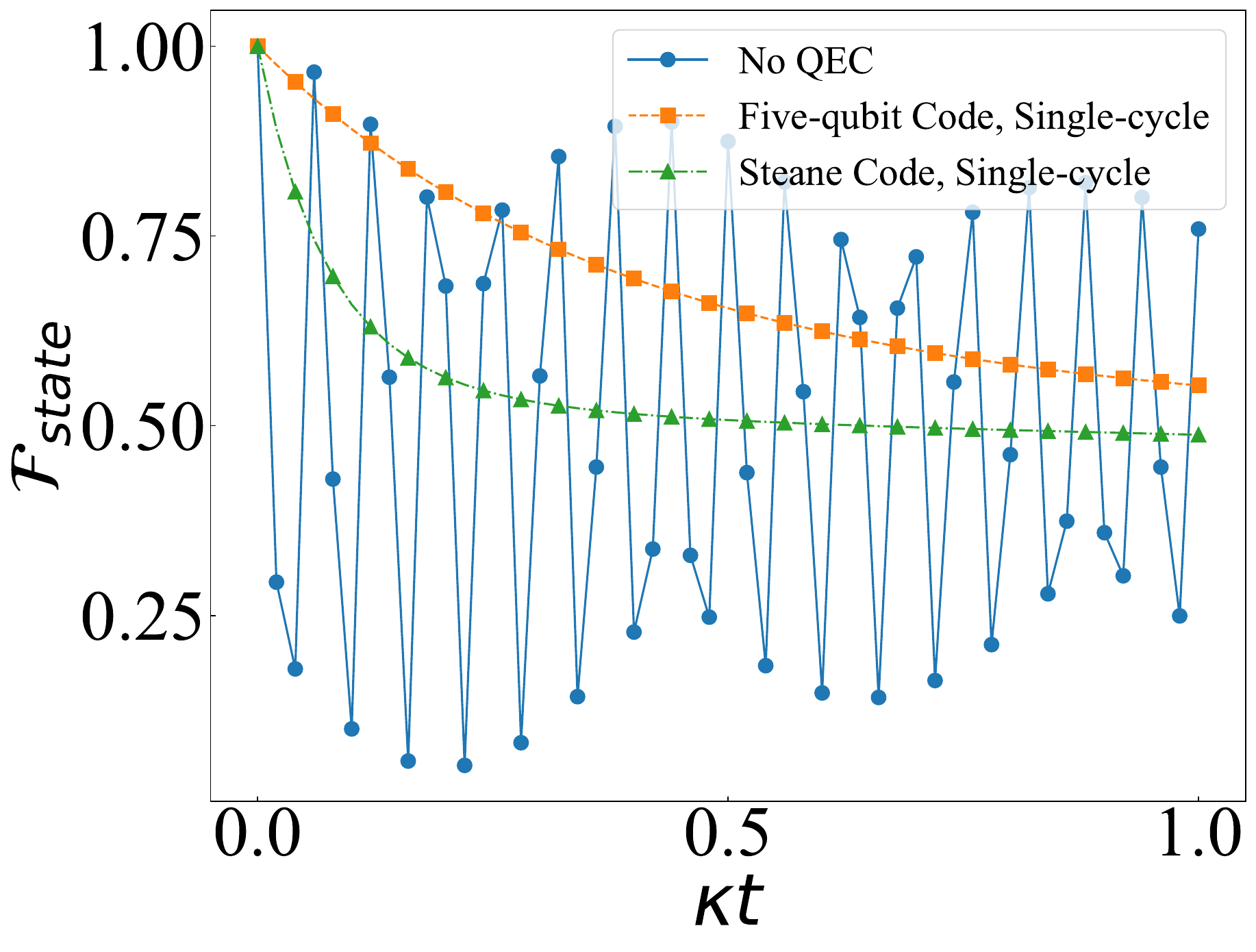}
\hfill
\subfigimg[width=0.49\textwidth]{b)}{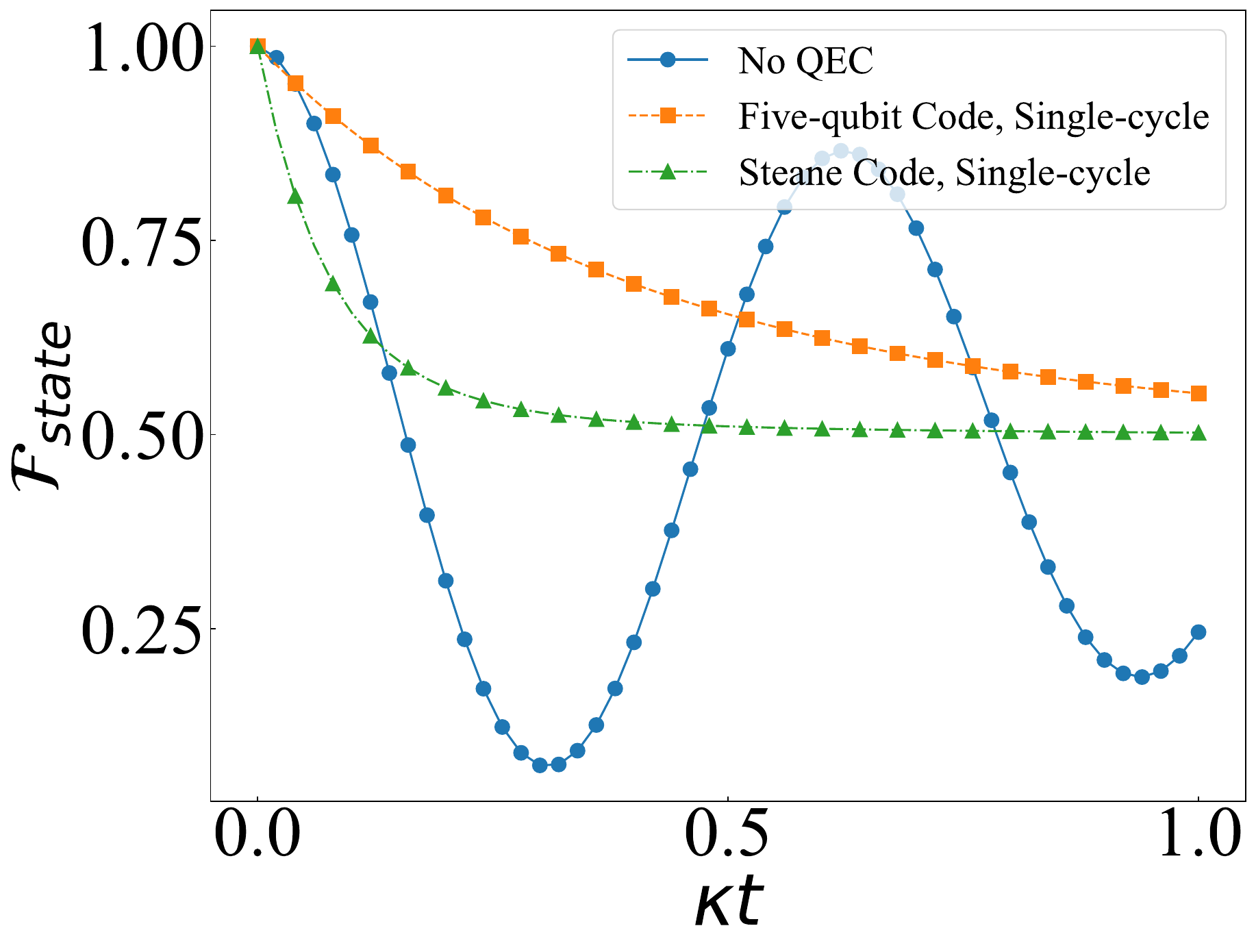}
\caption{State fidelity $\mathcal{F}_{state}$ of the initial state $\ket{+}$ with five-qubit QEC and Steane code. All qubits are coupled to baths with temperature $T=0.2$ for coupling strengths \idg{a} $\kappa/\omega=0.01$, and \idg{b} $\kappa/\omega=0.1$. The plot shows that the five-qubit code performs better than the Steane code.}
\label{fig11}
\end{figure*}

A comparative study of different QEC codes for the $\ket{+}$ initial state with different coupling strengths is shown in Fig.~\ref{fig11}.

\section{ Origin of the Critical Time for Werner States under QEC} 
\label{AppendixG}

Here, we provide a simple analytical argument to explain the emergence of a critical evolution time $t_c$ observed in the numerical results for two-qubit Werner states. This timescale marks the crossover beyond which QEC begins to improve the fidelity relative to the uncorrected evolution.

Let $\rho_0$ denote the initial state and $\rho(t)$ the evolved state. We compare the uncorrected evolution state $\rho_{\text{bare}}(t)$, with the QEC corrected evolution state $\rho_{\text{QEC}}(t)$.

Let the corresponding fidelities with respect to $\rho_0$ be
\begin{equation}
F_{\text{bare}}(t), \quad F_{\text{QEC}}(t).
\end{equation}
The critical time $t_c$ is defined as
\begin{equation}
F_{\text{QEC}}(t_c) = F_{\text{bare}}(t_c).
\end{equation}

From the master equation dynamics, the leading-order decay of fidelity for small $t$ can be approximated as
\begin{equation}\label{eqnF_bare}
F_{\text{bare}}(t) \approx 1 - \Gamma_{\text{eff}}\, t + \mathcal{O}(t^2),
\end{equation}
where $\Gamma_{\text{eff}}$ is an effective decoherence rate determined by the system--bath coupling and temperature
\begin{equation}
\Gamma_{\text{eff}} \sim J(\omega)\big(2n(\omega)+1\big).
\end{equation}

In a short time, the accumulated physical error is small, i.e.,
\begin{equation}
p(t) \sim \Gamma_{\text{eff}}\, t \ll 1.
\end{equation}
For a distance-3 code (such as the five-qubit code), QEC suppresses errors to leading order as
\begin{equation}
p_{\text{logical}} \sim \mathcal{O}(p^2).
\end{equation}

However, the application of QEC introduces an effective overhead, which we model as a small fidelity penalty $\delta$, arising from the projection onto the code space, mismatch between physical and logical error processes, and redistribution of errors across encoded qubits.
Thus, the corrected fidelity can be approximated as
\begin{equation}
F_{\text{QEC}}(t) \approx 1 - \delta - c\, p(t)^2,
\end{equation}
where $c$ is a constant depending on the code.
Substituting $p(t) \sim \Gamma_{\text{eff}} t$, we obtain
\begin{equation}\label{eqnF_QEC}
F_{\text{QEC}}(t) \approx 1 - \delta - c\, (\Gamma_{\text{eff}} t)^2.
\end{equation}

Now we know that the crossover condition is
\begin{equation}
F_{\text{QEC}}(t_c) = F_{\text{bare}}(t_c).
\end{equation}
Using \eqref{eqnF_bare}, \eqref{eqnF_QEC} we get
\begin{equation}
1 - \delta - c(\Gamma_{\text{eff}} t_c)^2 = 1 - \Gamma_{\text{eff}} t_c.
\end{equation}

Rearranging, we have
\begin{equation}
\Gamma_{\text{eff}} t_c = \delta + c(\Gamma_{\text{eff}} t_c)^2.
\end{equation}
For small $\delta$, the leading-order solution is
\begin{equation}
t_c \approx \frac{\delta}{\Gamma_{\text{eff}}}.
\end{equation}

Therefore, increasing temperature ($n(\omega)$) means larger $\Gamma_{\text{eff}}$ which results to smaller $t_c$.

\end{document}